\documentclass[a4paper, fleqn, usenatbib, useAMS]{mnras}

\usepackage[T1]{fontenc}
\usepackage{ae, aecompl}			
\usepackage{newtxtext, newtxmath} 	
\usepackage{mathtools}
\usepackage{graphicx}
\usepackage{amsmath}
\usepackage{amssymb}
\usepackage{multicol}
\usepackage{bm}
\usepackage{pdflscape}
\usepackage{natbib}
\usepackage[section]{placeins}
\usepackage{etoolbox}

\usepackage{xcolor}
\usepackage{geometry}

\newcommand{\psec}{$pc$}

\newcommand{\mega}{$M$}

\title[Secondary Spin Bias]{The Secondary Spin Bias of Dark Matter Haloes}

\author[J.W. Johnson et al.]{
	James W. Johnson,$^{1,2}$\thanks{Contact e-mail: \href{mailto:
	johnson.7419@osu.edu}{johnson.7419@osu.edu}} 
	Ariyeh H. Maller,$^{3,4}$ 
	Andreas A. Berlind,$^{2}$ 
	Manodeep Sinha,$^{2,6,7}$ \newauthor and 
	J. Kelly Holley-Bockelmann$^{2,5}$
	\\
	$^{1}$Department of Astronomy, The Ohio State University, 140 W. 18th Ave. 
	Columbus, OH 43210, USA \\ 
	$^{2}$Department of Physics and Astronomy, Vanderbilt University, 2301 
	Vanderbilt Place, Nashville, TN 37235, USA \\ 
	$^{3}$Department of Physics, New York City College of Technology, CUNY, 
	300 Jay St., Brooklyn, NY 11201, USA \\ 
	$^{4}$Department of Astrophysics, American Museum of Natural History, 
	New York, NY, USA \\ 
	$^{5}$Department of Physics, Fisk University, 1000 17th Ave. N, 
	Nashville, TN 37208, USA \\ 
	$^{6}$Centre for Astrophysics and Supercomputing, Swinburne University of 
	Technology, Hawthorn, Victoria 3122, Australia \\ 
	$^{7}$ARC Centre of Excellence for All Sky Astrophysics in 3 Dimensions 
	(ASTRO 3D)
}

\date{Accepted XXX; Received YYY; in original form ZZZ}
\pubyear{2018}

\begin{document}
\label{firstpage}
\pagerange{\pageref{firstpage}--\pageref{lastpage}}
\maketitle

\begin{abstract}
We investigate the role of angular momentum in the clustering of dark matter 
haloes.  We make use of data from two high-resolution N-body simulations 
spanning over four orders of magnitude in halo mass, from $10^{9.8}$ to 
$10^{14}\ h^{-1}\ \text{M}_\odot$. We explore the hypothesis that mass 
accretion in 
filamentary environments alters the angular momentum of a halo, thereby 
driving a correlation between the spin parameter $\lambda$ and the strength of clustering. 
However, we do not find evidence that the distribution of matter on large 
scales is related to the spin of haloes. We find that a halo's spin is 
correlated with its age, concentration, sphericity, and mass accretion rate. 
Removing these correlations strongly affects the strength of secondary spin 
bias at low halo masses. We also find that high spin haloes are slightly more 
likely to be found near another halo of comparable mass. These haloes that are 
found near a comparable mass neighbour - a \textit{twin} - are strongly 
spatially biased. We demonstrate that this \textit{twin bias}, along with the 
relationship between spin and mass accretion rates, statistically accounts for 
halo spin secondary bias. 
\end{abstract}


\section{Introduction}
\label{sec:intro}
High-resolution cosmological N-body simulations suggest that the spatial 
clustering of dark matter haloes is related to secondary properties other than 
halo mass \citep{Sheth2004,Gao2005,Wechsler2006,Gao2007,Wang2007,Li2008,
Faltenbacher2010,Lacerna2012,Lazeyras2017,Villarreal2017,Salcedo2018}. 
These findings are notable because a simple spherical collapse model 
\citep{Press1974} for halo formation predicts that spatial clustering should 
solely depend on halo mass \citep{Mo1996}. This is a phenomenon dubbed 
\textit{secondary bias} by \citet{Salcedo2018} because of the clustering 
dependance on halo properties in addition to halo mass.  This has also
been called \textit{assembly bias} \citep{Croton2007} in the past. However, 
high mass haloes do not exhibit a secondary bias with age, but do with 
concentration \citep{Mao2018,Salcedo2018}. Furthermore, \citet{Salcedo2018} 
showed that the secondary age and concentration biases can be accounted for 
by a correlation between these properties and the distance to a massive 
neighbour ($M_\text{n} \geq 10M_\text{h}$). This mechanism is less related to 
the assembly histories of the haloes than it is a direct statement about the 
mass of the nearby neighbour. For this reason, we adopt the term secondary 
bias. 
\par
Secondary bias is of interest because it has the potential to shed light on 
the physical mechanisms by which dark matter haloes collapse and evolve. It is 
also of interest because of the many statistical models connecting galaxy 
clustering statistics to that of their host dark matter haloes, such as the halo 
occupation distribution \citep[e.g.][]{Peacock2000, Scoccimarro2001, 
Berlind2002, Cooray2002, Berlind2003, Zu2015, Zu2016}, the conditional 
luminosity function \citep[e.g.][]{Yang2003, vandenBosch2003}, and subhalo 
abundance matching \citep[e.g.][]{Vale2004, Conroy2006}. These methods have 
usually operated under the 
assumption that the clustering of dark matter haloes is dependent only on mass, 
or a single parameter related to the mass\footnote{Note however that abundance 
matching models do account for any clustering bias from substructure by 
design.}. It may be tempting to assume that galaxies would inherit a secondary 
bias from the property-dependent clustering of their host haloes. Models where 
galaxy color is determined by halo age \citep{Hearin2013} have been 
successfully used to explain the color dependance of galaxy clustering 
\citep{Hearin2014,Hearin2015}. So far unexplored is the suggested connection 
between galaxy formation physics and halo spin \citep{Fall1980,Mo1998,
Somerville2008}, although there are those who argue against such a connection 
\citep{MallerDekel2002,Brook2011,Brook2012,Somerville2018}. If halo spin 
affects galaxy formation physics, then halo spin secondary bias would 
impact the galaxy-dark matter connection. Conversely, secondary spin bias 
would allow one to test whether galaxy formation models where galaxy size 
depends on halo spin are consistent with observational data. 
\par
The first studies to quantify halo secondary bias found that the clustering 
strength at fixed halo mass depends on the value of a wide range of secondary 
halo properties, from age to environment. In particular, haloes that are older 
\citep{Sheth2004, Gao2005, Wechsler2006, Gao2007, Wang2007, Li2008}, more 
tightly concentrated \citep{Wechsler2006, Gao2007, Faltenbacher2010, 
Lazeyras2017, Villarreal2017}, spherically shaped \citep{Faltenbacher2010}, 
have more substructure \citep{Wechsler2006, Gao2007}, or high spin 
\citep{Gao2007, Faltenbacher2010, Lacerna2012, Lazeyras2017, Villarreal2017} 
exhibit stronger clustering than their counterparts of similar 
mass. All of these biases, with the exception of spin, grow stronger for lower 
mass haloes. Recently, \citet{Lazeyras2017} showed that the mass accretion rates 
of haloes exhibit a similar phenomenon, with slowly accreting haloes exhibiting 
stronger clustering than their more quickly accreting counterparts. 
This bias is also strongest at low halo masses. In the case of spin, however, 
the bias is strongest at high masses. \par
Early interpretations placed the root 
cause of this bias on differences in the assembly histories of haloes; hence 
the term \textit{assembly bias} from \citet{Croton2007}. This was an 
attractive interpretation not only for its simplicity, but also because it 
would explain why the biases from halo properties related to the assembly 
histories of the haloes (e.g. concentration, shape, and amount of substructure) 
show a similar dependence on mass. However, some properties that are also 
directly related to the assembly history, such the time of the last major 
merger, do not exhibit assembly bias \citep{Li2008,Salcedo2018}. This suggests 
that assembly history may not be the only causal mechanism behind these 
clustering biases. Recently, \citet{Salcedo2018} showed that halo age and 
concentration are strongly correlated with the distance to a massive neighbour. 
That is, when one takes a subsample of old or highly concentrated haloes, they 
systematically select the ones that are closer in space to a much more massive 
system. This \textit{neighbour bias} accounts for most of the secondary bias as 
a function of halo mass associated with halo age and concentration. Halo spin, 
however, showed only a weak dependence on the distance to a massive neighbour, 
and a correction for this relationship did not account for a significant 
portion of the associated signal. These results, along with the fact that the 
secondary bias from halo spin is seen to have the opposite dependence on halo 
mass, comprise the strong evidence that it has differenct causal mechanisms 
than the secondary biases from age and concentration. \par
\citet{Lacerna2012} suggested that in filaments, the preferred directionality 
of mass accretion could spin up the halo. Since filament haloes are in overdense 
regions, this could result in a correlation between halo spin and clustering 
statistics. In this paper, we test this hypothesis, as well as explore other 
possible explanations of halo spin secondary bias. We describe our simulations 
and the methods we use to measure halo clustering in section 
\ref{sec:data_methods}. We provide an overview of secondary spin bias in 
section \ref{sec:background}. We explore the correlation between halo spin and 
the anisotropy of the large scale matter distribution in section 
\ref{sec:anisotropy}. In section \ref{sec:other_secondary_properties}, we 
detail correlations between spin and other halo properties. We demonstrate the 
role of \textit{twin bias} in causing secondary spin bias in section 
\ref{sec:twin_bias}. We end with discussion in section \ref{sec:conclusion}. 

\section{Data and Methods}
\label{sec:data_methods}

\begin{table*}
\caption{A summary of the cosmological parameters for the simulations.}
\def\arraystretch{1.2}
\begin{tabular*}{\textwidth}{c @{\extracolsep{\fill}} c c c c c c c c}
\hline 
Name & Box Size [$\mega\psec\ h^{-1}$] & Particle Mass [M$_\odot\ h^{-1}$] &
$h$ & $\Omega_\text{M}$ & $\Omega_\Lambda$ & $n$ & $\sigma_8$ & 
$N_\text{haloes}$ \\
\hline
Vishnu & 130 & $3.215\times10^7$ & 0.70 & 0.25 & 0.75 & 1.0 & 0.8 & 
706,060 \\ 
ConsueloHD & 420 & $1.87\times10^9$ & 0.681 & 0.302 & 0.698 & 0.96 & 0.828 & 
707,880 \\
SMDPL & 400 & $9.63\times10^7$ & 0.678 & 0.307 & 0.693 & 0.98 & 0.8228 & 
8,603,741 \\
\hline
\end{tabular*}
\label{tab:simparams}
\end{table*}

\subsection{The Simulations}
\label{sec:simulations}
We use two high-resolution dark matter only simulations to study the 
secondary bias associated with halo spin. The first is named \textit{Vishnu}, 
which has a box width of $130\ h^{-1}\ \text{Mpc}$ and a particle mass of 
$3.215\times10^7\ h^{-1}\ \text{M}_\odot$. Since the box size for 
\textit{Vishnu} is relatively modest, the impact of cosmic variance can be 
large. To suppress the effects of cosmic variance, we needed to ensure that 
the modes close to the fundamental mode are not significantly different from 
linear theory. We generated 100 realizations of the initial conditions
using 100 different random number seeds. The initial conditions
with the smallest root-mean-square deviation from the linear theory power 
spectrum forms the basis for the \textit{Vishnu} simulation. 
The second simulation is named 
\textit{ConsueloHD}, which has a box width of $420\ h^{-1}\ \text{Mpc}$ and 
a particle mass of $1.87\times10^9\ h^{-1}\ \text{M}_\odot$. The initial power 
spectrum for both simulations was calculated using \texttt{CAMB} 
\citep{Lewis2011}. The initial positions and velocities of the particles at 
redshift $z$ = 99 were 
then determined using the \texttt{2LPT} code \citep{Scoccimarro1997}. Both 
simulations evolved to $z$ = 0 using the \texttt{GADGET-2} N-body 
TreeSPH algorithm \citep{Springel2005} in a $\Lambda$CDM cosmology. Their 
sizes, particle masses, hubble parameters $h$, 
mass energy fractions $\Omega_M$, dark energy fractions $\Omega_\Lambda$, 
power spectrum normalizations $n_\text{s}$, and the mass density flucuations 
at 8 $h^{-1}$ Mpc $\sigma_8$ are presented in Table \ref{tab:simparams} along 
with the number of haloes. We use the \texttt{ROCKSTAR} and 
\texttt{Consistent-Trees} algorithms to find haloes \citep{Behroozi2013,
Behroozi2011}, adopting the virial definition of haloes with $\Delta_\mathrm{vir}$
set by \citet{bryan1998}. We set a minimum mass resolution limit of 200 
gravitationally bound particles for each halo, corresponding to a virial mass 
of $10^{9.8}\ h^{-1}\ \text{M}_\odot$ and $10^{11.6}\ h^{-1}\ \text{M}_\odot$ 
for Vishnu 
and ConsueloHD, respectively. In this paper we only consider distinct haloes 
that do not live inside larger haloes (i.e. hosts or central haloes). We do 
not include subhaloes in the sample. We thus obtain a dataset of 706,060 and 
707,880 haloes from Vishnu and ConsueloHD, respectively. By combining these two 
simulations, we obtain data spanning over four orders of magnitude in halo 
mass; from $10^{9.8}$ to $10^{14}\ h^{-1}\ \text{M}_\odot$. 
\par
In order to ensure that the differences in cosmology between Vishnu and 
ConsueloHD do not impact our measurements of halo spin secondary bias, we 
briefly employ data from the \textit{Small MultiDark Planck} (SMDPL) 
simulation \citep{Prada2012,Klypin2016}. The cosmological parameters for this 
simulation are also summarized in Table \ref{tab:simparams} along with 
Vishnu and ConsueloHD. We again set a minimum mass resolution of 200 
gravitationally bound particles ($M_\text{h} = 10^{10.3}\ h^{-1}\ 
\text{M}_\odot$) and remove all subhaloes. We thus obtain a dataset of 
8,603,471 
haloes from SMDPL. By using this simulation, we obtain data that spans the 
range in halo mass in which Vishnu and ConsueloHD overlap. Because SMDPL has 
cosmological parameters similar to ConsueloHD but different from Vishnu, 
this allows us to ensure that our measurements of halo spin secondary bias 
are independent of cosmology. 
\par
It has been shown that halo spin takes a large number of particles to converge 
in N-body simulations \citep{Onorbe2014,Benson2017}. With 200 particles, there 
are errors of order unity, and at least $4\times10^4$ particles are required 
for 10\% precision. These errors will shuffle high and low spin haloes, 
\textbf{reducing} the secondary bias measured. In general, we consider just 
splitting the sample into two bins, and thus for low particle number haloes, 
one should interpret the measured spin biases as lower limits. For higher 
particle number haloes, this effect becomes negligible. 

\subsection{Halo Properties}
\label{sec:props}
We make use of each of the following halo properties in this analysis as 
measured using the \texttt{ROCKSTAR} and \texttt{Consistent-Trees} codes.

\textbf{1. Halo Spin:} We adopt a definition of halo spin according to a 
dimensionless parameter defined in \citet{Bullock2001}:
\begin{equation}
\label{eq:bullock_lambda}
\lambda \equiv \frac{J}{\sqrt{2}MVR}
\end{equation}
where $J$ is the total angular momentum of the halo, $M$ is its virial mass, 
and $V$ and $R$ are the virial circular velocity and virial 
radius, respectively.
We note that the original and an alternative definition of halo spin is 
defined in \citet{Peebles1969} as:
\begin{equation}
\label{eq:peebles_lammda}
\lambda_P \equiv \frac{J\sqrt{\left|E\right|}}{GM^{5/2}}
\end{equation}
where $E$ is the total energy of the halo (a negative value if it is bound) 
and $G$ is Newton's gravitational constant. In this paper, we present results 
using the \citet{Bullock2001} definition, but we note that we have repeated 
our analysis with the \citet{Peebles1969} definition, $\lambda_P$, and found 
qualitatively similar results.

\textbf{2. Age:} To measure age, we use the redshift at which a halo 
accumulated half of its present-day mass. This quantity is hereafter referred 
to as the \textit{half mass redshift}, or $z_{1/2}$. This definition dictates 
that haloes which have lost mass will tend to be old, because half of their 
present day mass is less than half of their peak mass. 

\textbf{3. Concentration:} To measure the halo concentration, we use the 
method introduced in \citet{Klypin2011}. In this method, the maximum circular 
velocity is measured for each halo, and this is used to infer the halo's 
concentration under the assumption that it has an NFW density profile 
\citep{NFW}. This has been shown to give the same value as the traditional 
definition of concentration $c_\text{vir} = R_\text{vir}/r_\text{s}$ when the 
density profile is well fit by the NFW profile. However, it provides much more 
reasonable values when the density profile deviates significantly from NFW. 

\textbf{4. Shape:} We measure halo shape using the eigenvalues of the moment of 
inertia tensor applied to the particles within the halo as determined by 
\texttt{ROCKSTAR} (see \citet{Behroozi2013} or section \ref{sec:anisotropy} 
for details). We use the square root of the ratio of the smallest to largest 
eigenvalues of this tensor
as an indicator of the sphericity of the halo, where $(c/a)_\text{halo}$ = 1 
corresponds to a spherically symmetric distribution of matter. \footnote{
	We include the subscript on $(c/a)_\text{halo}$ because we employ a 
	similar notation in quantifying the anisotropy of the large scale matter 
	distribution. 
} 
This quantity does not contain information about whether the halo itself is 
oblate or prolate. In fact, \citet{Schneider2012} showed that the more 
detailed shapes of haloes is a distribution which varies with virial mass. We 
thus employ this measure as a general diagnostic for ellipticity. 

\textbf{5. Accretion Rate:} To measure halo mass accretion rates, we use the 
total mass accreted over the previous dynamical time of each halo. This value 
is determined by \texttt{Consistent-Trees}, and is denoted here by 
$\langle\dot{M}\rangle$, where the brackets denote that it is an average over 
the previous dynamical time.  Since the accretion rate strongly depends on 
halo mass we use a relative accretion $\langle\dot{M}\rangle / \langle\dot{M}
\rangle_\text{max}$ where $\langle\dot{M}\rangle_\text{max}$ is the maximum 
accretion rate of a halo in a given mass bin. This dimensionless parameter is 
then mass independent and can be used to compare the relative accretion rates 
of haloes across a range of halo masses.

\textbf{6. Redshift of Last Major Merger:} The most recent redshift where the 
halo had a merger with a mass ratio of at least 1:3 is denoted as 
$z_\text{lmm}$. Haloes that never have a merger of this ratio are given a value 
of $z_\text{lmm}$ when they first have enough particles to be counted as a 
halo. 

\subsection{Statistical Measures}
\label{sec:statmeasures}

\begin{figure*} 
\includegraphics[scale = 0.5]{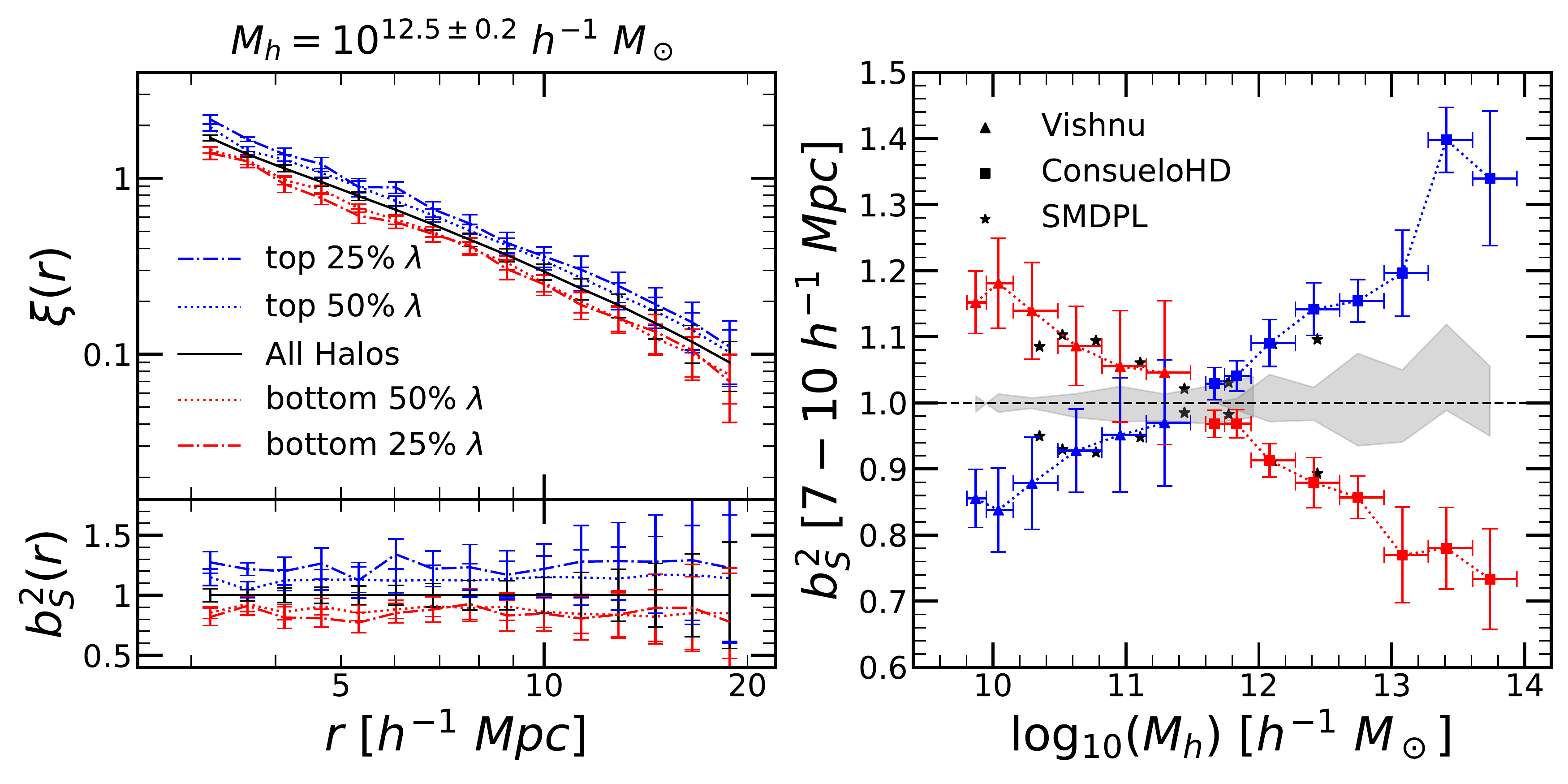}
\caption{
\textit{Top-Left:} The two-point autocorrelation as a function of scale range 
using only ConsueloHD haloes with virial masses between $10^{12.3}$ and 
$10^{12.7}\ h^{-1}\ \text{M}_\odot$. We show the autocorrelation of the entire 
mass range in the black solid line, and low and high spin samples in red and 
blue, respectively. 
\textit{Bottom-Left:} The ratio of the autocorrelations of each subsample and 
the entire mass bin. This is the relative bias for these subsamples as a 
function of scale range. 
\textit{Right:} The relative bias as a function of halo mass measured between 
7 and 10 $h^{-1}\ \text{Mpc}$ for the 50\% highest and lowest spinning haloes 
in each mass bin. The black dashed line corresponds to $b_S^2 = 1$, 
corresponding to the absence of a secondary bias by definition. The shaded 
region shows the finite bin width error, which is obtained by splitting bins 
in halo mass by the halo mass itself and measuring the relative bias. 
Triangles and squares denote measurements made with Vishnu and ConsueloHD 
haloes, respectively. We overplot in black stars the same measurements from the 
SMDPL simulation. The good agreement between Vishnu and SMDPL demonstrates 
that the small differences in cosmological parameters between these two 
simulations do not significantly impact secondary bias. 
}
\label{fig:figure1}
\end{figure*}

As a measure of halo clustering, we make use of the two-point autocorrelation 
function:
\begin{equation}
\label{eq:xi}
\xi(r) = \frac{DD(r)}{RR(r)} - 1
\end{equation}
where $DD(r)$ denotes the number of pairs of points separated within a scale 
range $r$ and $r + dr$ within a given subsample of data, and $RR(r)$ is the 
number of pairs we would expect if the points were distributed randomly. We 
note that this formulation of $\xi$ is generally noisier than that of 
\citet{Landy1993}. However, in this analysis, we take advantage of the 
simple geometry of a simulation box to calculate $RR(r)$ analytically, and 
therefore with no associated error:
\begin{equation}
RR(r) = \frac{1}{2}\frac{N^2}{L^3}\frac{4}{3}\pi\Big[(r + dr)^3 - r^3\Big]
\end{equation}
where $N$ denotes the number of haloes in the sample, and $L$ is the width of 
the box. By implementing the two-point autocorrelation in this manner, 
the only source of noise in our measurements is in $DD(r)$. Because of the 
large number of haloes in our box the associated error is small. 
We calculate these errors using the jackknife method. Haloes are split 
into eight subsamples corresponding to each of the eight octants of the box. 
$\xi(r)$ is then calculated 8 times, each with one octant removed from the 
data. The associated error in $\xi$ is then given by the jackknife formula 
with 8 subsamples:
\begin{equation}
\label{eq:jack_xi}
\sigma\xi(r) = \sqrt{\frac{7}{8}\sum_{i = 1}^{8}(\xi_{i}(r) - 
\langle\xi(r)\rangle)^{2}}
\end{equation}

As a measure of the secondary bias associated with some secondary property of 
a halo $S$, we retain the following definition from \citet{Salcedo2018}:
\begin{equation}
\label{eq:bsquared}
b_S^2(r|M_\text{h},S) = \frac{\xi_{hh}(r|M_\text{h},S)}{\xi_{hh}(r|M_\text{h})}
\end{equation}
where $\xi_{hh}(r|M_\text{h})$ denotes the autocorrelation of the haloes 
conditioned on some mass bin $M_\text{h}$. $\xi_{hh}(r|M_\text{h},S)$ is 
precisely that, but also conditioned on some halo secondary property $S$ (e.g. 
age, concentration, or 
spin). Under this formulation, $b_S^2 = 1$ for all haloes within a given mass 
range by design. \textbf{We emphasize that this is not the traditional 
definition of bias, which is relative to the underlying dark matter 
distribution.} This is a relative bias of a subsample against a halo mass bin 
which itself is biased compared to the dark matter density distribution. 
\par
We also use the jackknife method in determining errors in $b_S^2$. We again 
use the octants of the box as subsamples, and obtain errors in a 
manner analogous to that of equation \ref{eq:jack_xi}. That is, the 
autocorrelation and bias are separately calculated for each subsample with 
one octant removed. The variance is then calculated off of these eight 
measurements. However, the values for $\xi$ and $b_S^2$ that we report, are 
those which are obtained from all of the haloes. \par

\section{Halo Spin Secondary Bias}
\label{sec:background}

\begin{figure*} 
\centering
\includegraphics[scale = 0.33]{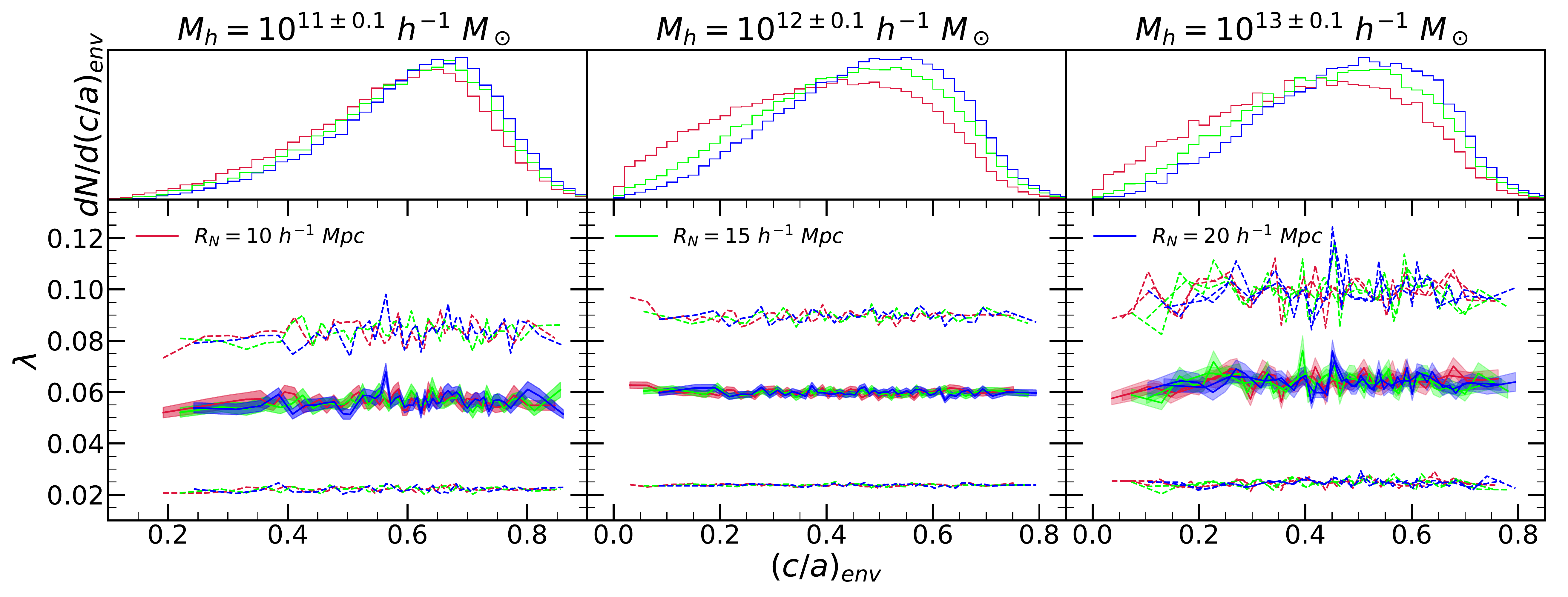}
\caption{
Measurements of the local anisotropy (i.e. non-sphericity) of the large-scale 
matter distribution as measured by the eigenvalues of the moment of inertia 
tensor applied to neighbouring haloes out to a distance $R_\text{N}$ from a 
central halo. We plot the implied 
minor-to-major axis ratio $(c/a)_\text{env}$ from this tensor against the 
spin parameter $\lambda$ for haloes with virial masses of 
$10^{11\pm0.1}$ (left), $10^{12\pm0.1}$ (middle), and $10^{13\pm0.1}\ h^{-1}\ 
\text{M}_\odot$ (right) for $R_\text{N}$ = 10 (red), 15 (green), and 20 
(blue) $h^{-1}\ \text{Mpc}$. Solid lines show the mean $(c/a)_\text{env}$ 
vs. $\lambda$ in equal number bins. Shaded regions correspond to the 
uncertainty in the mean spin, and the dashed lines denote the range in $\lambda$ 
containing 68.2\% of the haloes in that bin, centered on the median. Our 
$(c/a)_\text{env}$ statistic is a proxy for filamentary structure, and the 
distributions of haloes in this quantity for each value of $R_\text{N}$ are 
shown in the top panels. We use Vishnu haloes in the $10^{11\pm0.1}\ h^{-1}\ 
\text{M}_\odot$ sample, and ConsueloHD haloes in the remaining two. 
}
\label{fig:anisotropy}
\end{figure*}

We begin by measuring the scale and mass dependence of halo spin secondary 
bias via the statistical measures outlined in section \ref{sec:statmeasures}. 
The top-left panel of Figure \ref{fig:figure1} shows the two-point 
autocorrelation function for only ConsueloHD haloes with masses between 
$10^{12.3}$ and $10^{12.7}\ h^{-1}\ \text{M}_\odot$ for various subsamples in 
$\lambda$. Red and blue lines denote low versus high $\lambda$ subsamples, 
respectively, with dotted and dot-dashed lines 
differentiating between halves and quartiles. We show this in comparison to 
what is obtained when we take into account the entire mass range, shown by the 
solid black line. This is the essence of halo spin secondary bias. In a sample 
of haloes that have roughly the same mass, those that have higher values of 
$\lambda$ exhibit stronger clustering than what would be expected from their 
mass alone. In turn, those that are of lower $\lambda$ exhibit weaker 
clustering. 
\par
The bottom-left panel of Figure \ref{fig:figure1} shows the ratios 
that are obtained when we divide the autocorrelations of the spin subsamples 
by that of the entire mass range.  The color and line-style scheme is 
the same as discribed above. This ratio is the secondary bias, $b_S^2$. The 
secondary bias associated with $\lambda$ is relatively scale independent, 
consistent with a straight line within the errors. For the rest of this paper 
we will focus on the clustering in the range between 7 and 10 Mpc $h^{-1}$. 
Because of this scale independence, this decision is arbitrary and should not 
impact our results. We have verified that this is indeed the case. \par
While the secondary bias of other halo properties like age and concentration 
has been shown to be driven by the $\sim$10\% extremes of the distribution, 
this is not the case for spin \citep{Salcedo2018}. We have verified this 
result as well. The relatively linear increase in $b_S^2$ between the half 
and quartile subsamples in $\lambda$ also illustrates this. Because the 
increase in bias when using quartiles is rather modest, throughout this paper 
we split mass bins into halves in order to have more haloes in each bin. 
\par
The right-hand panel of Figure \ref{fig:figure1} shows the secondary bias at 
this length scale for a wide range of halo masses and simulations. We use 
triangles and squares to denote measurements made with Vishnu and ConsueloHD 
datasets, respectively. We also include $b_S^2$ as calculated from the SMDPL 
simulation \citep{Prada2012,Klypin2016}. We plot these measurements as black 
stars for both high and low $\lambda$ samples and omit the error bars for 
clarity. The shaded region denotes the relative bias that we 
measure when we split our bins in halo mass by the halo masses themselves, 
referred to as the \textit{finite bin width error} \citep{Salcedo2018}. This 
region quantifies the maximum error due to the intrinsic width of the mass bins used 
in our analysis, and represents the relative bias that would be measured 
from any secondary property if it were perfectly correlated with halo mass.
\par
As one can see, the measured relative bias is in excellent agreement across 
all three simulations. Since SMDPL has almost the same cosmological 
parameters to ConsueloHD, but different from Vishnu, we conclude that the 
relative spin bias is independent of cosmology, at least for the modest 
cosmology differences probed here.
For the remainder of this paper we use Vishnu and ConsueloHD. 
\par
Here we see results similar to that obtained by previous studies of halo 
spin secondary bias \citep{Gao2007, Faltenbacher2010, Lacerna2012, 
Salcedo2018}. Halo spin exhibits the strongest secondary bias in the high mass 
regime, and this falls off in lower mass ranges. We also see that this trend 
continues, producing a turnover at ${\sim}10^{11.5}\ h^{-1}\ \text{M}_\odot$, 
below which low $\lambda$ haloes exhibit slightly stronger clustering. This was 
also found recently by \citet{SatoPolito2018} using the SMDPL simulation. It 
appears that this feature has not been seen in previous studies of halo 
secondary bias because \citet{SatoPolito2018} and ours are the first studies 
to use simulations with a high enough resolution to make measurements at 
sufficiently low halo masses.
\par
We do note from Figure \ref{fig:figure1} that the relative bias decreases for 
the lowest mass bins in Vishnu and SMDPL. It is likely that these measurements 
are affected by the errors in spin associated with low particle number haloes 
\citep{Onorbe2014,Benson2017}, which will inevitably cause shuffling between 
high and low spin samples. This suggests that these are systematic 
underestimates of the true relative bias in these mass ranges. These 
measurements should therefore be interpreted qualitatively as lower limits on 
the relative bias. We however note that this source of error has no obvious 
effect outside of the lowest mass bin. 

\section{Local Anisotropy}
\label{sec:anisotropy}

\begin{figure*} 
\includegraphics[scale = 0.28]{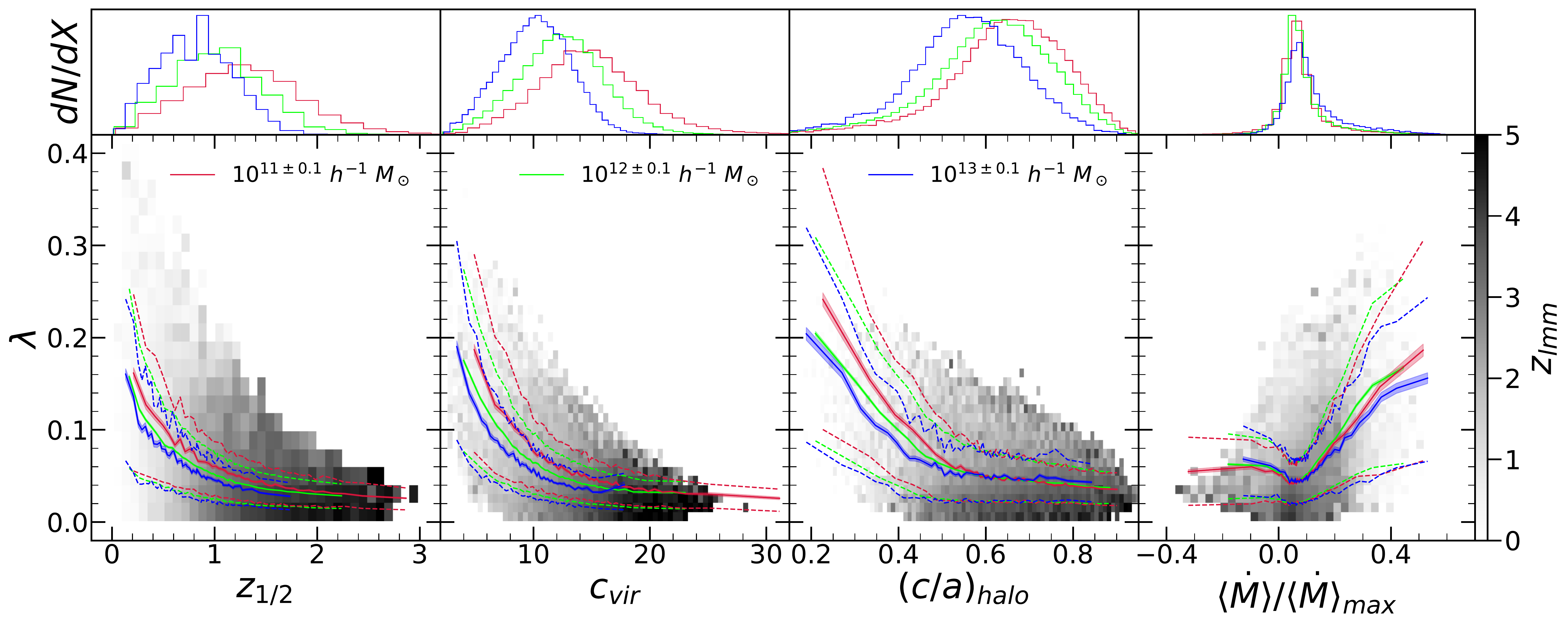}
\caption{
Halo spin $\lambda$ as a function of $z_{1/2}$ (left), $c_\text{vir}$ 
(left-middle), $(c/a)_\text{halo}$ (right-middle), and 
$\langle\dot{M}\rangle / \langle\dot{M}\rangle_\text{max}$ (right) for haloes 
with virial masses of $10^{11\pm0.1}$ (red), $10^{12\pm0.1}$ (green), 
and $10^{13\pm0.1}$ (blue) $h^{-1}\ \text{M}_\odot$. Solid lines show the mean 
spin as a function of the each parameter while the shaded regions show the error 
in that mean. Dashed lines denote the region of $\lambda$ containing 68.2\% of 
the haloes in each bin, centered on the median. Shown in grayscale is the mean 
redshift of the last major merger $z_\text{lmm}$ within each pixel, where a merger 
event is counted if the mass ratio of the merging haloes is at least 1:3. 
Top panels show the probability distribution functions of haloes in each 
secondary property for each of the three mass ranges. The $10^{11\pm0.1}$ 
$h^{-1}\ \text{M}_\odot$ sample contains haloes from Vishnu, while the 
remaining two contain haloes from ConsueloHD.
}
\label{fig:correlations}
\end{figure*}

We continue by testing the hypothesis raised by \citet{Lacerna2012} that halo 
spin secondary bias is driven by filamentary structure. They conjectured that 
dark matter accretes onto haloes with a preferred directionality in 
filaments. This could cause an increase in halo angular momenta, and thereby 
spin. Filamentary environments are more likely to be overdense and could 
therefore connect $\lambda$ to clustering statistics. To study this, we 
measure the local anisotropy (i.e. non-sphericity, elongation) of the halo 
distribution of surrounding haloes in the following manner. 
For each halo above our resolution limit, we measure the moment of inertia 
tensor of the haloes in the local neighbourhood defined by a 
\textit{neighbourhood radius}, denoted $R_\text{N}$. For a neighbourhood with 
$N$ haloes having masses denoted by $m_i$ at a separation of 
$d_i \leq R_\text{N}$ from the host, the moment 
of inertia tensor is given by:
\begin{equation}
\label{eq:tensor}
I \equiv \sum_{i = 1}^{N}m_i\left[
\begin{matrix}
dy_{i}^2 + dz_{i}^2 & -dx_{i}dy_{i} & -dx_{i}dz_{i} \\ 
-dx_{i}dy_{i} & dx_{i}^2 + dz_{i}^2 & -dy_{i}dz_{i} \\ 
-dx_{i}dz_{i} & -dy_{i}dz_{i} & dx_{i}^2 + dy_{i}^2
\end{matrix}
\right]
\end{equation}
This tensor has three positive definite eigenvalues. We denote the square root 
of the ratio of the smallest to largest eigenvalue as $(c/a)_\text{env}$, 
analogous to the axis ratios for haloes as discussed in section 
\ref{sec:props}. \par
In Figure \ref{fig:anisotropy}, we present results comparing halo spin to 
$(c/a)_\text{env}$ in three different mass ranges ($M_\text{h} \approx 10^{11}$ 
(red), $10^{12}$ (green), and $10^{13}\ h^{-1}\ \text{M}_\odot$ (blue)) and 
for three different neighbourhood radii ($R_\text{N} = 10$ (left), 15 (middle), 
and 20 $h^{-1}$ Mpc (right)). The lowest mass bin 
contains haloes only from Vishnu, while the other two use haloes from ConsueloHD. 
The upper panels show the normalized distributions of 
haloes in $(c/a)_\text{env}$. They clearly show that with increasing 
$R_\text{N}$, 
the distribution shifts gradually toward values closer to unity. This is as 
expected, since on larger scales we expect the matter distribution to become 
more isotropic until it eventually reaches a fully homogenous distribution. 
\par
The lower panels of Figure \ref{fig:anisotropy} show the mean $\lambda$ in 
equal number bins in $(c/a)_\text{env}$ in each mass range and for each 
neighbourhood radius. Shaded regions denote the associated error in the mean 
$\lambda$, and dotted lines quantify the 68.2\% dispersion in $\lambda$ for 
each bin, centered on the median. There does not appear to be any significant 
correlation between halo spin and $(c/a)_\text{env}$ in any mass range or on 
any scale. That is, $(c/a)_\text{env}$ appears to uniformly sample the 
distribution of haloes in $\lambda$. \par
While we have only presented correlations between $\lambda$ and 
$(c/a)_\text{env}$ here, we note that we have found similar results with 
$(b/a)_\text{env}$ and $(c/b)_\text{env}$. We also find that 
$(c/b)_\text{env} = 0.9 \pm 0.1$ for nearly all haloes in both Vishnu and 
ConsueloHD. This suggests that a prolate spheroid is a reasonable description 
of the mass distribution. Our $(c/a)_\text{env}$ variable is thus a proxy 
for filamentary structure. 
\par
It appears that $\lambda$ is not statistically sensitive to filamentary 
structure. This would suggest that filamentary environments do not produce 
systematically high spin haloes, and without this correlation, they cannot be 
the physical explanation for its associated secondary bias. Whether or not 
$(c/a)_\text{env}$ shows its own secondary bias would be irrelevant, because 
these environments would uniformly sample the spin distribution. 

\section{Age, Concentration, Sphericity, and Mass Accretion}
\label{sec:other_secondary_properties}

\begin{figure*} 
\includegraphics[scale = 0.28]{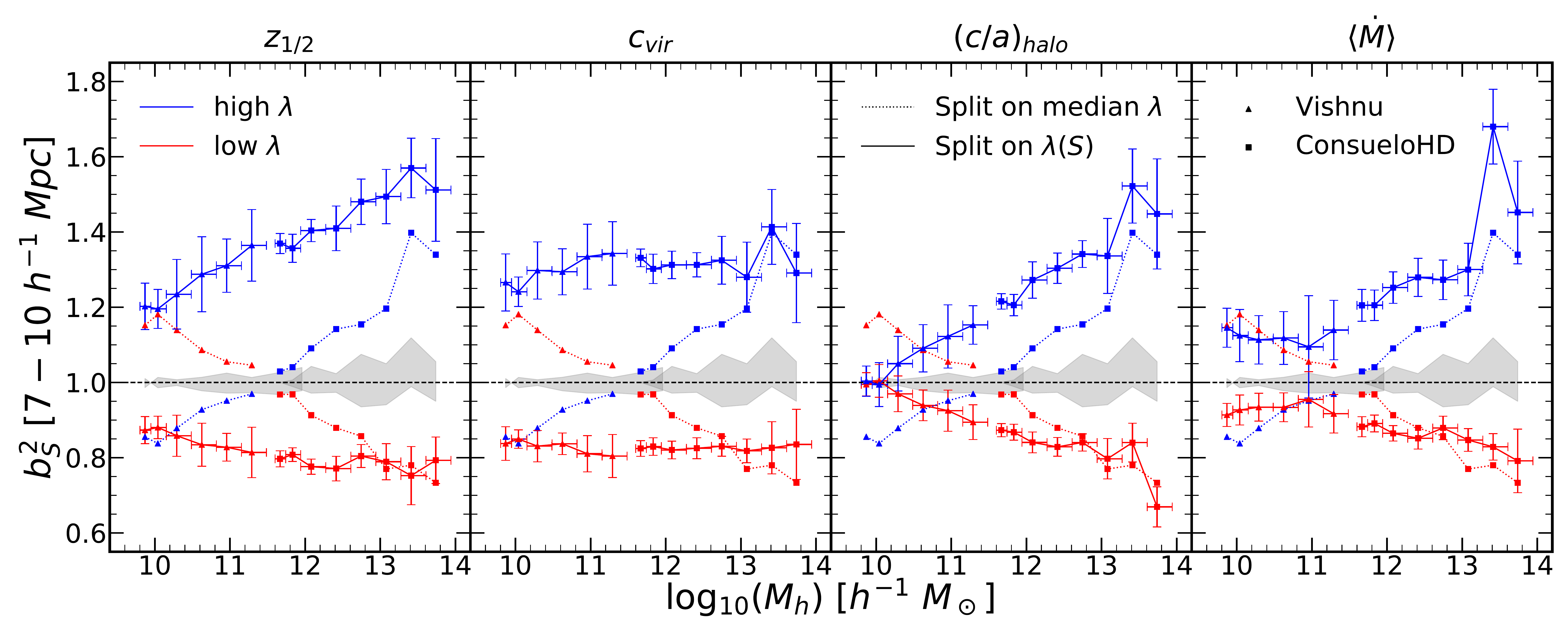}
\caption{
The relative bias as a function of halo mass for high (blue) and low (red) 
spin subsamples. Dotted lines show the relative bias that we measure when 
splitting mass bins into the 50\% highest and lowest spinning haloes (these
lines are identical to the dotted lines in the right panel of Fig.~\ref{fig:figure1}).
Solid lines show the relative bias that we measure when we draw high and low 
spin samples while controlling for $z_{1/2}$ (left panel), $c_\text{vir}$ 
(left-middle panel), $(c/a)_\text{halo}$ (right-middle panel), and 
$\langle\dot{M}\rangle$ (right panel). We do this by fitting a sixth-order 
polynomial to the median $\lambda$ as a function of each property, and selecting
haloes with spins above and below this curve. This is done within each individual 
bin in halo mass. Triangles and squares represent measurements made with Vishnu and 
ConsueloHD haloes, respectively. The shaded region denotes the finite bin width 
error, which we obtain by splitting bins in halo mass by the halo masses 
themselves and measuring the relative bias. The black dashed line highlights 
$b_S^2 = 1$. 
}
\label{fig:bias_zhalf_c_ctoa_mdot}
\end{figure*}

Finding that the anisotropy of the large scale halo distribution is not 
correlated with halo spin and therefore not likely to be responsible for its 
associated secondary 
bias, we turn to examining the correlation between spin and other halo 
properties in search of an explanation for this phenomenon. 
In this section we present results showing correlations between halo spin and 
halo age, halo concentration, halo sphericity, and the rate of mass accretion 
onto the halo as we have defined them in Section \ref{sec:props}. 
Figure \ref{fig:correlations} 
shows these relationships for the $10^{11\pm0.1}$, $10^{12\pm0.1}$, and 
$10^{13\pm0.1}\ h^{-1}\ \text{M}_\odot$ mass ranges in red, green, and blue, 
respectively. As in Figure \ref{fig:anisotropy}, we plot the 
mean $\lambda$ in equal number bins in each secondary property, with shaded 
regions denoting the error in the mean and dashed lines 
quantifying the 68.2\% variance centered on the median in each bin. In the 
right-hand panel, we show the normalized mass accretion rate where the 
accretion rate is divided by the maximum rate in that mass bin. The one object 
with a value of unity in this quantity is off the plot as all other values, 
by construction, are less than this. 
\par
In the top panels, we show the normalized distributions of haloes in each of 
these secondary properties. These distributions illustrate known trends of 
these secondary properties with halo mass. Higher mass haloes tend to be 
younger, less concentrated, and more ellipsoidal than those of lower 
mass. When the mass accretion rates of haloes are normalized in this manner, 
however, it appears to be rather uniform across mass ranges.
We note that there are strong inverse relationships between $\lambda$ and 
each of $z_{1/2}$, $c_\text{vir}$, and $(c/a)_\text{halo}$. Thus high 
$\lambda$ haloes tend to be younger, less concentrated, and more aspherical 
than other haloes. We also note a strong direct relationship between $\lambda$ 
and $\langle\dot{M}\rangle$; high $\lambda$ haloes tend to also be accreting 
mass at a faster rate. \par 
To understand why these correlations exist, we overplot in gray scale the mean 
redshift at which haloes in the mass range $10^{12\pm0.1}\ h^{-1}\ 
\text{M}_\odot$ 
have had a major merger (mass ratio greater than 1:3). Clearly high spin haloes 
are more likely to have had recent major mergers, which is not surprising 
because in a major merger the orbital angular momentum of the haloes is 
converted to spin angular momentum \citep{Vitvitska2002,Maller2002,Drakos2018a,
Drakos2018b}. 
Furthermore, age, concentration, sphericity, and mass accretion all also show 
correlation with the redshift of the last major merger. Younger haloes will 
tend to have lower $z_\text{lmm}$ because our measurement of halo age 
$z_{1/2}$ is directly related to its mass assembly history. The relationship 
between $z_\text{lmm}$ and $c_\text{vir}$ is probably due to mergers; 
following a merger event, the core of the halo will require a finite amount of 
time to dynamically relax. This will increase the scale radius of the halo, 
thus decreasing the concentration. As new material accretes onto the halo, not 
though major mergers, it will add mass to the outer part of the halo, 
increasing the virial 
radius and thus increasing the concentration. Likewise a major merger will 
tend to enhance the density in the plane of the merger, making the halo's shape 
more ellipsoidal. Finally, after a major merger, it is not surprising to find a 
higher accretion rate. This is not only because major mergers are often 
associated with increased rates of accretion, but also because some material 
in the merger may take multiple orbits before being included in the virial 
mass. 
\par
We note that the grayscale in the $\lambda$-$c_\text{vir}$ plane extends to 
$z_\text{lmm} \approx 5$. In the simulated cosmology, redshifts this high 
correspond to lookback times comparable to the Hubble time. This is much 
larger than the dynamical time of any physically realistic halo, indicating 
that the response to a merger event is not the only physical 
process influencing the spins of these haloes. This suggests that something 
related to a lack of major mergers is also influencing this correlation. 
\par

\begin{figure*} 
\includegraphics[scale = 0.33]{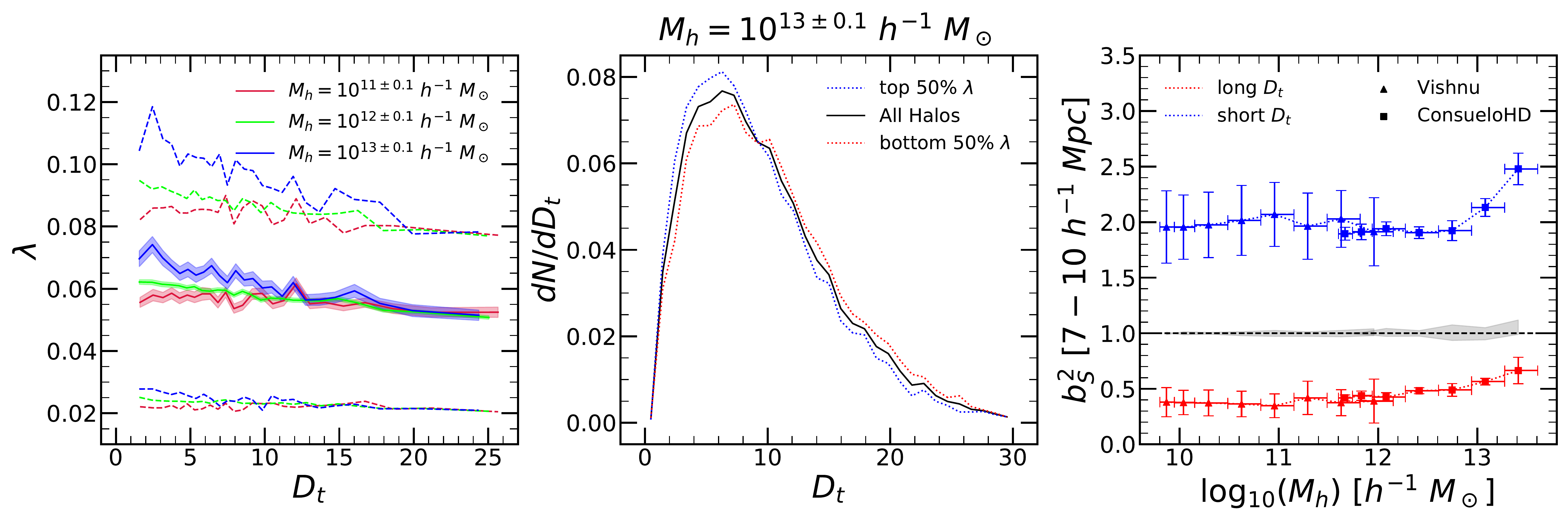}
\caption{
\textit{Left:} Spin vs. twin distance for haloes with virial masses of 
$10^{11\pm0.1}$ (red), $10^{12\pm0.1}$ (green), and $10^{13\pm0.1}$ (blue) 
$h^{-1}\ \text{M}_\odot$. Twin distance refers to the distance from a halo 
to a neighbouring halo between 1 and 3 times its own mass divided by the 
neighbour's virial radius. We plot the mean halo spin in equal number bins of 
$D_t$, with shaded regions denoting the error in that mean. Dashed 
lines denote the range in $\lambda$ containing 68.2\% of 
haloes, centered on the median $\lambda$ in that bin. The $10^{11\pm0.1}$ 
$h^{-1}\ \text{M}_\odot$ sample contains haloes from Vishnu, while the 
remaining two contain haloes from ConsueloHD. 
\textit{Middle:} Normalized distributions of $D_t$ for the 
$10^{13\pm0.1}\ h^{-1}\ \text{M}_\odot$ sample from ConsueloHD. The black 
solid line shows the distribution for the entire mass range, while colored 
lines show that of the spin subsamples. 
\textit{Right:} The relative bias as a function of halo mass for the 50\% 
shortest (blue) and longest (red) $D_t$ samples. Triangles and squares 
denote measurements made from Vishnu and ConsueloHD haloes, respectively. 
The black dashed line represents $b_S^2 = 1$. The shaded region 
shows the finite bin width error, which we obtain by splitting our bins in 
halo mass by the halo mass itself and measuring the relative bias. 
}
\label{fig:dtwin}
\end{figure*}

Each of the secondary properties plotted against spin in Figure 
\ref{fig:correlations} has an associated secondary bias. Specifically, the 
secondary bias associated with each of $z_{1/2}$, $c_\text{vir}$, 
$(c/a)_\text{halo}$, and $\langle\dot{M}\rangle$ is such that haloes that tend 
to be low $\lambda$ exhibit stronger clustering than those that 
tend to be high $\lambda$ in the low mass regime. This means that one would 
naively expect low spin haloes to exhibit stronger clustering, which is not 
true for haloes with masses $\text{M}_\text{h} \gtrsim 
10^{11.5}\ h^{-1}\ \text{M}_\odot$. To decouple spin from these correlated 
properties, we adopt the following statistical correction.
For each mass bin within which we measure $b_S^2$, we first fit a sixth-order 
polynomial to the median relationship between $\lambda$ and one of these secondary 
properties $S$. We then draw high and low $\lambda$ subsamples based off of 
the haloes that lie above or below the curve that we obtain. Thus the split is 
done for $\lambda(S)$, the spin as a function of another parameter, rather 
then the median lambda in each mass bin $\tilde{\lambda}$. This procudure
ensures that our high and low spin samples have identical distributions in S.
\par
After drawing high and low $\lambda$ samples in this manner for each of 
$z_{1/2}$, $c_\text{vir}$, $(c/a)_\text{halo}$, and $\langle\dot{M}\rangle$, 
we measure the relative bias as a function of halo mass, and plot our results 
in Figure \ref{fig:bias_zhalf_c_ctoa_mdot}. From left to right, the panels 
show the secondary bias associated with $\lambda$ as a function of $z_{1/2}$, 
$c_\text{vir}$, $(c/a)_\text{halo}$, and $\langle\dot{M}\rangle$. Solid lines 
show the values of $b_S^2$ for spin split based on a function of another halo 
property $\lambda(S)$, while 
dotted lines show the relative bias when split by the median spin in each bin. 
The dotted lines here are the same as the dotted lines in Figure 
\ref{fig:figure1}; for clarity, we have omitted their error bars. The 
clustering of high and low $\lambda$ haloes is shown in blue and red, 
respectively. Because the adjusted subsamples are drawn from a polynomial fit, 
they do not necessarily split the subsamples evenly. 
\par
When splitting the samples based on spin as a function of another halo 
property, we see similar behavior across all four parameters studied here. 
Most notably, the turnover in the signal previously seen 
at $10^{11.5}\ h^{-1}\ \text{M}_\odot$ 
is no longer present. In all cases, the slope of the halo mass 
dependence of the secondary bias is reduced; for concentration, it becomes 
nearly flat. That is, we find that removing spin's correlation with age, 
concentration, halo shape, or mass accretion rate removes much of the halo 
mass dependence of $b_S^2$, while still producing a secondary bias. 
Thus we see that in the low mass regime, secondary spin bias is 
closely intertwined with age, concentration, halo shape, and mass accretion 
rate. As these four properties are all correlated with one another, it is 
likely that we are seeing a single phenomenon responsible for most of the halo 
mass dependence for spin secondary bias. 
We note that at the high mass end, the effect of splitting on spin as a 
function of another halo property is much smaller. In the next section, we 
explore an environmental halo property to explain this residual secondary 
bias. 

\section{Twin Bias}
\label{sec:twin_bias}

\begin{figure*}  
\includegraphics[scale = 0.47]{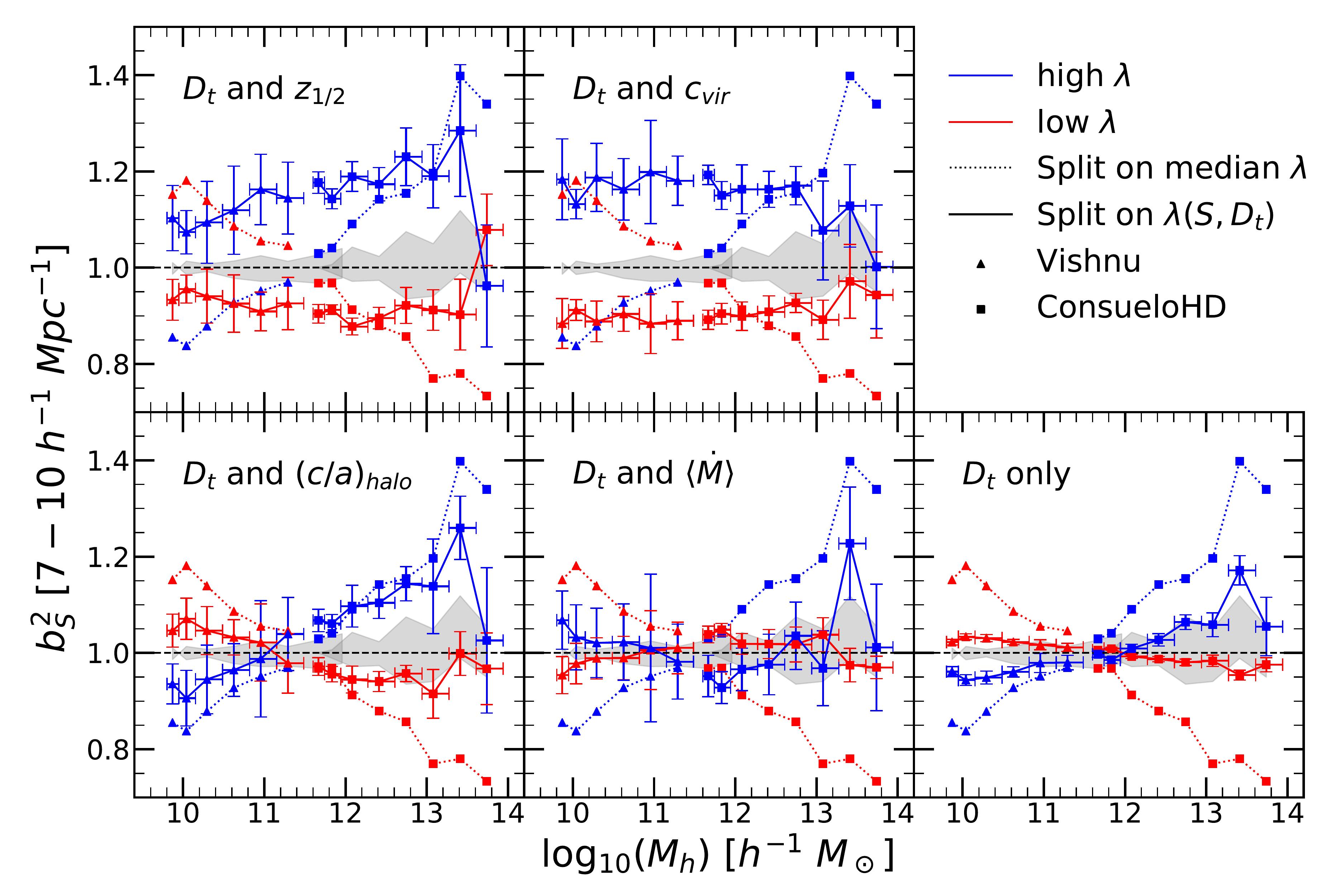}
\caption{
The relative bias as a function of halo mass. Dotted lines show that of the 
50\% highest and lowest spinning haloes in each mass bin (and are identical 
to the dotted lines in the right panel of Fig.~\ref{fig:figure1}). Solid 
lines show the relative bias for spin samples that are obtained while 
controlling for one or two secondary properties in each individual mass bin 
(i.e. such that both spin samples have statistiscally similar distributions 
in these properties). In the bottom-right panel, this is done for twin 
distance $D_t$ only. This refers to the distance to a neighbouring halo 
between 1 and 3 times as massive as the halo itself divided by that neighbour's 
virial radius. In the remaining panels, we jointly control for both $D_t$ and 
each of $z_{1/2}$ (top-left), $c_\text{vir}$ (top-middle), $(c/a)_\text{halo}$ 
(bottom-left), and $\langle\dot{M}\rangle$ (bottom-middle). The black dashed 
line represents $b_S^2 = 1$ everywhere, while the shaded region denotes the 
finite bin width error, which is obtained when we split our bins in halo mass 
by the mass itself and measure the relative bias. 
} 
\label{fig:bias_dt_all}
\end{figure*}

Based on the results of the previous section, it appears that none of the 
environmental or halo properties we have explored thus far can account for 
secondary spin bias. In this section, we therefore turn to studying a new 
environmental property: a halo's distance to a neighbouring halo of comparable 
mass. \citet{Salcedo2018} showed that there was little correlation between 
spin and the distance to a ten or greater times as massive halo. Here we find 
that there is a correlation between spin and the distance to a halo of 
comparable mass, or a \textit{twin} halo. We adopt the following unitless 
halo distance parameter from \citet{Salcedo2018}: 
\begin{equation}
D_{t} = \frac{D}{R_\text{vir,t}}
\label{eq:dtwin}
\end{equation}
where $D$ denotes the halo-centric distance from the host to the neighbour of 
comparable mass, and $R_\text{vir,t}$ is the \textbf{neighbour's} virial radius. 
We have changed subscripts from $D_{n}$ (adopted by \citealt{Salcedo2018}) 
to $D_{t}$ to emphasize that this 
distance is to a twin in mass - not a much more massive neighbour. 
For every halo, we record the distance to the nearest twin, where the twin is 
required to be a halo of mass between 1 and 3 times the halo's mass. 
We plot $\lambda$ of haloes in three mass ranges against this quantity in the 
left-hand panel of Figure \ref{fig:dtwin}. Results are shown for haloes of 
virial masses $10^{11\pm0.1}$, $10^{12\pm0.1}$, and $10^{13\pm0.1}$ 
$h^{-1}\ \text{M}_\odot$ in red, green, and blue, respectively. Shaded 
regions here denote the error in the mean $\lambda$ in equal number bins, 
while dotted lines show the range in $\lambda$ containing 68.2\% of haloes, 
centered on the median value. The lowest mass bin contains haloes from Vishnu, 
while the other two bins are composed of haloes from ConsueloHD.
\par 
Figure \ref{fig:dtwin} shows a rather weak, yet clear anticorrelation between 
$\lambda$ and twin-distance which gets stronger with increasing halo mass. 
This relationship does not flatten off, but appears to continue to large 
values of $D_{t}$. At high $D_t$, 
haloes in all mass ranges have a mean spin of $\lambda \approx 0.05$, but in 
the highest mass range, the lowest $D_t$ haloes have a mean spin of $\lambda 
\approx 0.07$. Highly elongated or non-concentrated haloes, however, have a 
mean spin of $\lambda \approx 0.2$, a much more substantial effect.
\par
In the middle panel, we show probability distribution functions in $D_{t}$ 
for haloes in the $10^{13\pm0.1}\ h^{-1}\ \text{M}_\odot$ sample. The black line 
denotes the entire data in this mass range, while colored lines denote high 
and low samples in $\lambda$. These distributions clearly show a selection 
effect associated with $\lambda$, whereby subsamples of high $\lambda$ haloes 
are systematically more likely to be found at low $D_{t}$ - that is, fewer 
virial radii from a neighbour of comparable mass. In this same mass range, the 
high spin sample contains 22 and 14\% more haloes at $D_t \leq$ 5 and 10 than 
the low spin sample, respectively. 
\par
While the relationship between $\lambda$ and $D_t$ is subtle, the 
distance to a twin is a very spatially biased property, as can be seen in the 
right-hand panel of Figure \ref{fig:dtwin}. This panel shows a strong 
secondary bias of $b_S^2 \approx 2$ for low $D_t$ haloes at lower masses, which 
then begins to increase near the redshift dependent collapse mass 
$M_* \approx 3\times10^{12}\ h^{-1}\ \text{M}_\odot$. Because the 
$\lambda$-$D_t$ relation and the intrinsic spatial bias of twin haloes are both 
mass-dependent, it is likely that this influences the mass dependence of 
secondary spin bias for high mass haloes. 
\par
This bias is significantly stronger than the secondary spin bias, 
as well as the secondary biases from age, concentration, halo shape, and mass 
accretion. This is unsurprising, because low $D_t$ haloes are by 
definition near a neighbour of comparable mass. Despite the subtlety of spin's 
relationship with twin distance, it's secondary bias may be significantly 
sensitive to $D_t$ due to this strong spatial bias. 
\par
We test the sensitivity of halo spin secondary bias to this relationship by 
fitting a sixth-order polynomial under the same prescription as in section 
\ref{sec:other_secondary_properties} to $\lambda$ as a function of $D_{t}$. 
We assign haloes to be high or low $\lambda$ based on whether or not they lie 
above or below this curve. We show our results in the bottom-right panel of 
Figure \ref{fig:bias_dt_all}. In this figure, the dotted lines are the same as 
in Figure \ref{fig:figure1}, the bias when $\lambda$ is split by the median in 
each mass bin; we have again omitted the error bars on these measurements for 
visual clarity. Solid lines show the relative bias that is measured when high 
and low $\lambda$ subsamples in each mass bin are split by spin as a 
function of $D_{t}$, thus ensuring that they have identical $D_{t}$ 
distributions. Under this prescription, the relative bias decreases 
significantly in all mass ranges. Above the turnover at $10^{11.5}\ h^{-1}\ 
\text{M}_\odot$, the signal that remains is largely consistent with 
the finite bin width error to within $1\sigma$, with an outlier in one mass 
bin. Below the turnover, while its strength diminishes substantially, a small 
relative bias remains. It would thus appear that the 
subtle relationship between $\lambda$ and $D_t$ is enough to statistically 
account for a large portion of the signal associated with secondary spin bias. 
\par
For the other panels in Figure \ref{fig:bias_dt_all}, we determine spin as a 
function of both twin distance and a second halo property. In each mass bin, 
we fit a 2-dimensional polynomial surface to $\lambda$ as a function of both 
$D_t$ and another secondary property. The function fit is simply the sum of 
two sixth-order polynomials: one a function of $D_t$ and the other a function 
of $z_{1/2}$, $c_\text{vir}$, $(c/a)_\text{halo}$ or $\langle\dot{M}\rangle$. 
We then assign haloes to be high or low spin based on the fitted spin value, 
$\lambda(D_t, S)$. In other words, we control for both $D_{t}$ and one of 
these secondary properties simultaneously.
\par
The top-left and top-middle panels show $b_S^2$ as a function of the log of 
the halo mass in $h^{-1}\ \text{M}_\odot$ when we correct our high and low 
$\lambda$ subsamples for $D_{t}$ in combination with $z_{1/2}$ and 
$c_\text{vir}$, 
respectively. The secondary bias in these panels is now close to mass 
independent. That is, the combination of spin's dependence on twin distance 
and either age or concentration seems to account for the mass dependance that 
we see in spin secondary bias when split on the median in each mass bin. For 
the highest mass haloes studied here, our measurements are within the finite 
bin width error, and therefore our results would be 
consistent with the interpretation that this prescription statistically 
accounts for secondary spin bias at these masses. 
\par
The bottom-left panel of Figure \ref{fig:bias_dt_all} shows the signal that is 
measured when high and low $\lambda$ subsamples are fit for both $D_{t}$ 
and $(c/a)_\text{halo}$. This panel demonstrates that this statistical 
correction only partially accounts for the relative bias. In the low-mass 
regime, the relative bias is reduced to about half of its original value. 
In the high-mass regime, a substantial portion of the low $\lambda$ 
under-clustering is accounted for, while the high $\lambda$ over-clustering 
remains largely unaffected. That is again with the exception of the highest 
mass bin. 
\par
The bottom-middle panel of Figure \ref{fig:bias_dt_all} shows the secondary 
bias when $\lambda$ is a function of both $\langle\dot{M}\rangle$ and $D_{t}$. 
This panel does not show a statistically significant secondary bias. 
Thus it seems to be the case that spin secondary bias can be accounted for by 
the mass accretion rate onto haloes and distance to a twin halo. That is, haloes 
that are found at similar $D_t$ with similar mass accretion rates do not 
exhibit secondary spin bias.  While some of 
the points are still greater than the finite bin width error, our fits of spin 
as a function of twin distance and mass accretion rate have some associated 
error which is not included here. Since all of these properties are correlated 
with one another, it is difficult and potentially misleading to identify one 
of age, concentration, sphericity, or mass accretion as being solely 
responsible for secondary spin bias beyond the contribution of $D_t$. 

\section{Conclusion}
\label{sec:conclusion}

We used haloes from three high resolution N-body simulations spanning from 
$10^{9.8}$ to $10^{14}\ h^{-1}\ \text{M}_\odot$ in virial mass. 
In section \ref{sec:background}, we measured the strength of halo spin 
secondary bias in all three of these simulations. Our measurements 
demonstrated that the measured 
relative bias in each mass bin is consistent across all three simulations. 
This suggests that the signal associated with halo spin secondary bias is 
independent of the cosmological parameters of the simulation. We now 
summarize the primary results of our work here. 

\begin{itemize}
\item We confirmed the finding of \citet{SatoPolito2018} that for haloes with 
virial masses of $M_\text{h} \lesssim 10^{11.5}\ h^{-1}\ \text{M}_\odot$, 
the secondary spin bias signal inverts. 
\item In section \ref{sec:anisotropy}, we presented measurements of the local 
anisotropy surrounding haloes in both Vishnu and ConsueloHD. We found that it 
does not correlate with halo spin, and thus the \citet{Lacerna2012} 
explanation for secondary spin bias is not supported. 
\item In section \ref{sec:other_secondary_properties}, we noted strong inverse 
correlations between spin and each of halo age, concentration, and 
sphericity, implying that high spin haloes tend to be young, 
non-concentrated, and obliquely shaped. Previous studies 
have found evidence suggesting that secondary spin bias has different causal 
mechanisms than that of age and concentration 
\citep{Gao2007, Faltenbacher2010, Lacerna2012, Salcedo2018}. These 
anticorrelations suggest that the old and highly concentrated haloes which 
exhibit strong clustering are a different sample than the high spin haloes 
which also exhibit strong clustering. That is, these secondary biases are 
produced by two unique populations, which confirms the previous studies' 
interpretations. 
\item All of the properties studied in section 
\ref{sec:other_secondary_properties} 
correlate with the redshift of the last major merger and show 
similar secondary bias behavior which is likely driven by neighbour bias 
\citep{Salcedo2018}. While it has been noted before that halo spin tends to 
increase in a merger event \citep{Maller2002,Vitvitska2002,Drakos2018a,
Drakos2018b}, this effect 
extends over the entire redshift range we have probed. This suggests that the 
lack of a major merger for a long time is just as important in determining a 
halo's spin. 
\item We calculated the dimensionless distance 
$D_t$ to neighbouring haloes of comparable mass ($\text{M}_\text{h} \leq 
\text{M}_\text{n} \leq 3\text{M}_\text{h}$), and found a slight, yet 
statistically significant anticorrelation between the halo spin $\lambda$ and 
$D_t$. The slope of this correlation is larger for higher mass haloes. This 
suggests that high spin haloes are statistically more likely to be found near a 
neighbouring halo of comparable mass - a \textit{twin}. This relationship 
introduces a selection effect which we refer to as \textit{twin bias}, whereby 
high spin haloes are slightly more likely to be found near a twin halo. This 
selection effect is such that in the $10^{13\pm0.1}\ h^{-1}\ \text{M}_\odot$ 
mass regime, splitting haloes into the top and bottom 50\% spin samples results 
in the high spin sample containing $\sim$15\% more $D_t \lesssim$ 10 haloes 
than the low spin sample. Furthermore, the quantity $D_t$ exhibits a strong 
spatial bias, which is unsurprising since low $D_t$ haloes are by definition 
near another halo of comparable mass. 
\item While the $\lambda$-$D_t$ relation is weak, we demonstrated that when 
high and low spin samples are drawn from a high-order polynomial fit to 
$\lambda$ as a function of $D_t$, a large portion of the secondary bias signal 
is removed. The only statistically significant signal that remained was below 
the turnover at $10^{11.5}\ h^{-1}\ \text{M}_\odot$. This suggests that twin 
bias plays a key role in producing the spin dependence of halo clustering. 
Furthermore, we also saw that when we account for twin bias along with halo 
age or concentration, a mild, mass-independent secondary bias is produced. 
When we account for twin bias in combination with halo sphericity, the 
relative bias is reduced but a significant signal remains. 
In the case of mass accretion rate, it nearly disappears entirely, with 
outliers in only a few mass bins. That is, haloes with similar mass accretion 
rates found at similar distances to comparable mass neighbours do not exhibit 
secondary spin bias. However, because all of these properties are related to 
one another, it is difficult and potentially misleading to identify one of 
age, concentration, sphericity, or mass accretion as the primary driver of 
this residual signal. We therefore conclude that secondary spin bias is a 
two-term phenomenon, where the signal is largely driven by twin bias, but is 
also influenced by the correlation between spin and other halo 
properties which also exhibit secondary biases. 
\par
\end{itemize}
\citet{Salcedo2018} showed that the same formulation of distance applied to 
more massive neighbours ($\text{M}_\text{n} \geq 10\text{M}_\text{h}$) does 
not show a correlation with spin. Twin bias only applies to neighbouring haloes 
of comparable mass, and for this reason it is fundamentally different from 
neighbour bias. When a halo is near a much more massive neighbour 
($\text{M}_\text{n} \gtrsim 3\text{M}_\text{h}$), this causes arrested 
development, which prevents high angular momentum material from accreting onto 
the halo. When the neighbour has less mass ($\text{M}_\text{n} \sim 
\text{M}_\text{h}$), the twin tidally torques the halo, but does not affect 
accretion since it is not heavy enough to cause arrested development. 
\par
There are a couple possible interpretations of the relationship between $\lambda$ 
and $D_{t}$ that we note here. First, it could be that the 
probability of a halo having a twin correlates with the large scale tidal 
field. Second, simply the presence of the twin may produce a tidal 
torque. The first interpretation, but not the second, would explain why the 
correlation between $\lambda$ and $D_t$ appears to continue at large $D_t$. 
However, these interpretations are not mutually exclusive, and further studies 
will be required to explore the full details of twin bias. 

\section{Acknowledgements}
J.W.J. would like to thank Robert Scherrer and Jonathan Bird at Vanderbilt 
University for participating in a review committee with J.K.H-B. and A.A.B. 
when this project was at the undergraduate thesis level. Some of the 
computation for this work was done using the Advanced Computing Center for 
Research and Education (ACCRE) at Vanderbilt University. J.W.J. also 
acknowledges the valuable comments on this work contributed by David H. 
Weinberg and Andres N. Salcedo at The Ohio State University. A.A.B. was
supported by the National Science Foundation (NSF) through a Career Award 
(AST-1151650).
This work used the Extreme Science and Engineering Discovery Environment
(XSEDE)~\citep{XSEDE14}, which is supported by National Science Foundation 
grant number ACI-1548562, through allocations TG-AST080002N and TG-AST130037.
MS would like to thank Matt Becker for sharing the memory-efficient
\texttt{uber-LGadget2} that was used to run the ConsueloHD and Vishnu 
simulations. Parts of this research were conducted by the Australian Research 
Council Centre of Excellence for All Sky Astrophysics in 3 Dimensions (ASTRO 
3D), through project number CE170100013. The authors gratefully acknowledge 
the Gauss Centre for Supercomputing e.V. (www.gauss-centre.eu) and the 
Partnership for Advanced Supercomputing in Europe (PRACE, www.prace-ri.eu) 
for funding the MultiDark simulation project by providing computing time on 
the GCS Supercomputer SuperMUC at Leibniz Supercomputing Centre (LRZ, 
www.lrz.de).

\bibliographystyle{mnras}
\bibliography{draft5}

\bsp
\label{lastpage}
\end{document}